\documentclass[a4paper,twocolumn,superscriptaddress,11pt,accepted=2018-07-05, groupedaddress] {quantumarticle}
\pdfoutput=1
\usepackage{times, mathrsfs, amsmath, amsfonts, graphics, graphicx, cancel, color, theorem, bbm, mathtools, amssymb}

\usepackage[english]{babel}
\usepackage[margin=.75in]{geometry}
\usepackage[T1]{fontenc}
\usepackage[numbers,sort&compress]{natbib}

\usepackage{xfrac,relsize,float}

\usepackage[unicode=true, breaklinks=false, pdfborder={0 0 1}, backref=false, colorlinks=true, linkcolor=blue, citecolor=blue]{hyperref}

\def\ket#1{\mathinner{|{#1}\rangle}}

\def\BraVert{\egroup\,\mid\,\bgroup}

\def\ketbra#1#2{|#1\rangle \!\langle#2|}
\definecolor{Blue}{rgb}{0,0,1}
\definecolor{Red}{rgb}{1,0,0}
\definecolor{Green}{rgb}{0,1,0}
\definecolor{darkgreen}{rgb}{0,.7,0}
\definecolor{Purp}{rgb}{.2,0,.2}
\definecolor{white}{rgb}{1,1,1}

\newcommand{\tr}{{\rm tr}}
\newcommand{\tre}{{\rm tr}_\etxt}
\newcommand{\trs}{{\rm tr}_\stxt}

\newcommand{\rmd}{{\rm d}}

\newcommand{\s}{{\scriptscriptstyle{S}}}
\newcommand{\e}{{\scriptscriptstyle{E}}}

\newcommand{\se}{{\scriptscriptstyle{SE}}}

\newcommand{\Mod}[1]{\,\mathrm{mod}\, #1}

\newcommand{\stxt}{S}
\newcommand{\etxt}{E}
\newcommand{\setxt}{SE}

\newcommand{\dg}{\dagger}

\begin{document}

\title{Tomographically reconstructed master equations for any open quantum dynamics}

\author{Felix A. Pollock}
\email{felix.pollock@monash.edu}
\orcid{0000-0002-1483-5661}
\affiliation{School of Physics \& Astronomy, Monash University, Clayton, Victoria 3800, Australia}
\author{Kavan Modi}
\email{kavan.modi@monash.edu}
\orcid{0000-0002-2054-9901}
\affiliation{School of Physics \& Astronomy, Monash University, Clayton, Victoria 3800, Australia}

\date{\today}
\begin{abstract}
Memory effects in open quantum dynamics are often incorporated in the equation of motion through a superoperator known as the memory kernel, which encodes how past states affect future dynamics. However, the usual prescription for determining the memory kernel requires information about the underlying system-environment dynamics. Here, by deriving the transfer tensor method from first principles, we show how a memory kernel master equation, for any quantum process, can be entirely expressed in terms of a family of completely positive dynamical maps. These can be reconstructed through quantum process tomography on the system alone, either experimentally or numerically, and the resulting equation of motion is equivalent to a generalised Nakajima-Zwanzig equation. For experimental settings, we give a full prescription for the reconstruction procedure, rendering the memory kernel operational. When simulation of an open system is the goal, we show how our procedure yields a considerable advantage for numerically calculating dynamics, even when the system is arbitrarily periodically (or transiently) driven or initially correlated with its environment. Namely, we show that the long time dynamics can be efficiently obtained from a set of reconstructed maps over a much shorter time.
\end{abstract}
\maketitle

%%%%%%%%%%%%%%%%%%%%%%%%%%%%%%%%%%%%
%%%%%%%%%%%%%%%%%%%%%%%%%%%%%%%%%%%%
\section{Introduction}
%%%%%%%%%%%%%%%%%%%%%%%%%%%%%%%%%%%%
%%%%%%%%%%%%%%%%%%%%%%%%%%%%%%%%%%%%

A large fraction of active research in physics and chemistry, both theoretical and experimental, involves characterising or modelling the dynamics of open quantum systems, usually in terms of the time evolution of the system's density operator. Such dynamics is, in general, significantly more complicated than that dictated by Schr\"odinger's equation. Nevertheless, many techniques have been developed to predict how open systems will evolve in time. These include schemes for tomographically determining an unknown open process~\cite{Chuangqpt, Modi2012} (even going beyond density operator evolution to the full multi-time statistics of observables~\cite{nonMarkovPRA,nonMarkovPRL, partialprocesstensor}), perturbative approaches to theoretically modelling dynamics~\cite{breuerpetruccione, Fruchtman2016, Iles-Smith2016}, and computationally expensive numerical techniques which can exactly simulate an open system under certain circumstances~\cite{heom, heom2, dmrg, quapi,devega2017}. However, the resources required for the latter often scale exponentially with the evolution time, and numerical shortcuts usually rely on assumptions about the dynamics, such as no time-dependent driving~\cite{CerrilloCao2014} or limited memory effects~\cite{Nalbach2011, BrendonQUAPI, TEMPO}, meaning long-time dynamical simulations are intractable for the most general open processes. Here, among other results, we develop a method for extracting the long-time dynamics for general driven open quantum systems.

\begin{figure}[thp]
\centering
\includegraphics[width=0.46\textwidth]
{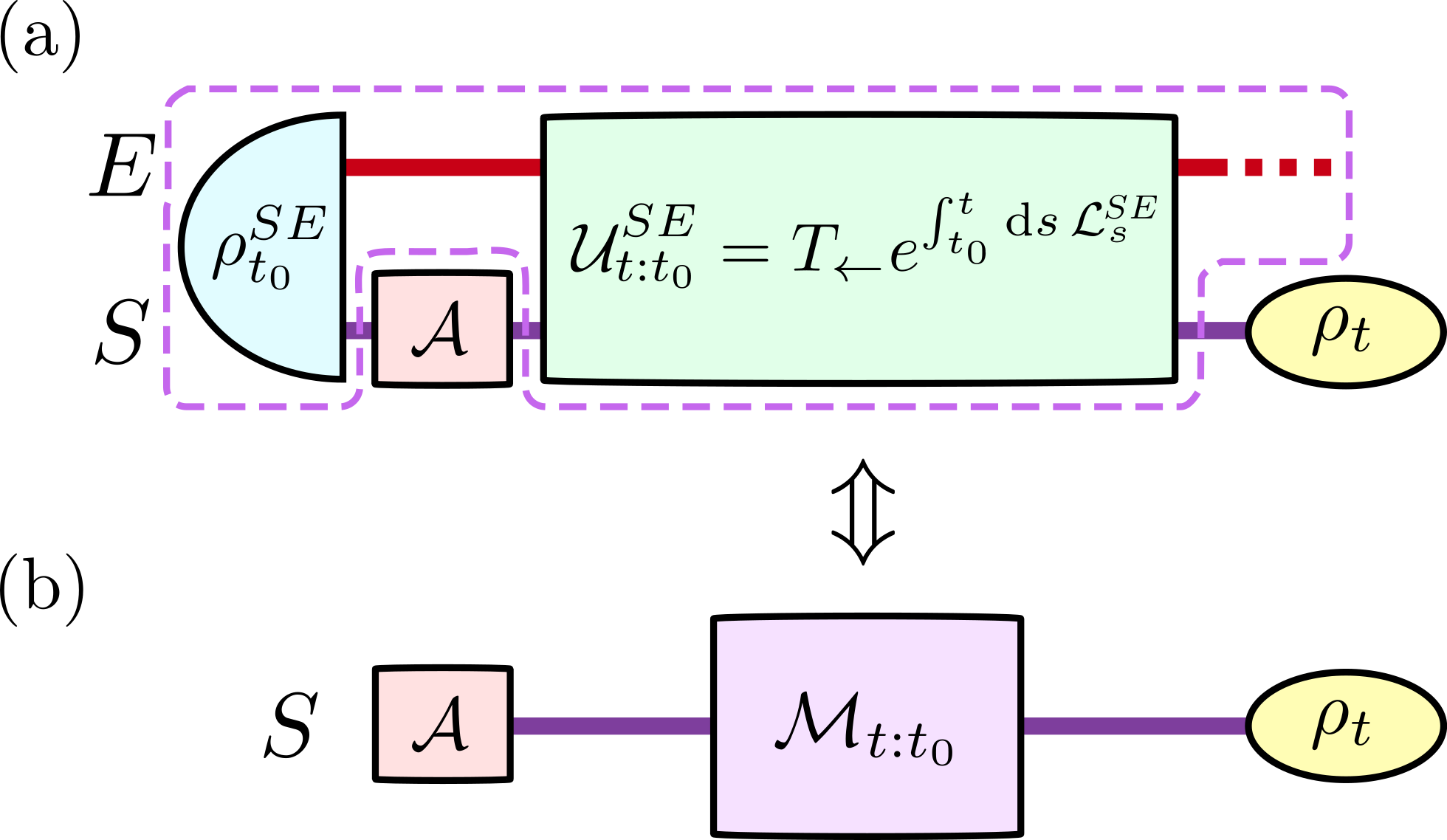}
\caption{\textbf{Two equivalent pictures of open quantum dynamics.} (a) The system, initially in joint state $\rho^{\se}_{t_0}$ with its environment, is prepared at time $t_0$ with operation $\mathcal{A}$. The two then evolve together according to $\dot{\rho}^{\se}_{t} = \mathcal{L}^\se_t \rho^{\se}_{t}$, leading to a final state $\rho_{t}$ for the system that depends on the initial preparation. (b) The system state at time $t$ can also be thought of as a function of the preparation operation at time $t_0$. All the uncontrollable parts of the process (enclosed within the dashed lines in the first picture), including the initial system-environment state, are encapsulated in the linear map $\mathcal{M}_{t:t_0}$. When $\rho^{\se}_{t_0}$ is uncorrelated, the dynamics can be equivalently thought of as a completely-positive trace-preserving map from the initial state of the system.}
\label{fig:twopictures}
\end{figure}

Conceptually, approaches to describing open quantum dynamics tend to take one of two perspectives, depicted in Fig.~\ref{fig:twopictures}: First is the ``dilated'' or ``underlying'' picture, where the system in question ($\stxt$) is represented as evolving with its environment ($\etxt$) from an initial state $\rho^\se_{t_0}$, with a time evolution superoperator $\mathcal{U}^\se_{t:t_0}$\footnote{Superoperator $\mathcal{U}^\se_{t:t_0}$ need not be unitary in general; however, it must be divisible.} that propagates solutions to an equation of the form 
\begin{gather}
\dot{\rho}^\se_t=\mathcal{L}^\se_t\rho^\se_t. \label{eq:underlying}
\end{gather} 
All the physics of the system is encoded in the superoperator $\mathcal{L}^\se_t$, which is usually taken to evolve $\setxt$ according to the von Neumann equation\footnote{That is, $\mathcal{L}^\se_t\rho^\se_t = -\frac{i}{\hbar}[H^\se_t,\rho^\se_t]$, where $H^\se_t$ is the $\setxt$ Hamiltonian.}, but could also incorporate dissipative and decohering dynamics through (possibly time-dependent) Lindblad terms~\cite{Lindbladeqn,GKSeqn}. 

The second picture is the operational one, where the entire dynamics is described in terms of quantities that would be, in principle, directly measurable by an experimenter with access to the system alone. Such a description contains only that information necessary to predict the evolution of the system, nothing more, and is usually more compact than a corresponding description in the dilated picture. In this picture, the dynamics may often be described in terms of a dynamical map $\Lambda_{t:t_0}$, which is completely positive and trace preserving, and evolves the initial system state $\rho_{t_0}$ to its counterpart at time $t$: $\rho_t = \Lambda_{t:t_0} \rho_{t_0}$\footnote{Throughout this Article we denote operators and superoperators acting on the system-environment space with superscript $SE$. We do not put a superscript on operators and superoperators acting on the system alone.}. This description is valid when the system and environment are initially uncorrelated in the other picture. More generally, the dynamics can be described in terms of a superchannel~\cite{modiosid,Modi2012} $\mathcal{M}_{t:t_0}$, which maps an initial preparation superoperator $\mathcal{A}$---any transformation that could be applied to the system at time $t_0$---to the state at later times: $\rho_t=\rho_{t,\mathcal{A}} = \mathcal{M}_{t:t_0} [\mathcal{A}]$. The superchannel guarantees the complete positivity of the dynamics even in the presence of initial correlations with an environment. The time-evolved system state that one would obtain without performing any preparation procedure (the `freely-evolved' state) is simply given by $\mathcal{M}_{t:t_0} [\mathcal{I}]$, where $\mathcal{I}$ is the identity superoperator. The dynamics, $\Lambda_{t:t_0}$ or $\mathcal{M}_{t:t_0}$, can be reconstructed operationally by means of quantum process tomography~\cite{Chuangqpt, martin, nonMarkovPRA}.

While these two pictures are equivalent, it is not always clear how to convert from the operational to the dilated picture (the reverse is straightforwardly achieved by tracing out over the environment, $\rho_t=\tre\rho^\se_t$).  
That is, recovering the underlying $\setxt$ quantities, $\rho^\se_{t_0}$ and $\mathcal{L}^\se_t$, from operationally reconstructible time-dependent maps, such as $\mathcal{M}_{t:t_0}$, is neither uniquely constrained nor easy to achieve in practice; though, in the case of smooth dynamical maps, the underlying objects have been shown to be continuous and smooth themselves~\cite{Kretschmann2008,Dive2015}. Where such a recovery is possible, even partially, the system can be used as a probe of its environment, extracting information about its Hamiltonian directly~\cite{NorrisPazSilva2016, probingPRA}, or otherwise determining the parameters of a phenomenological master equation~\cite{Jeske2012, Bellomo2009, Bellomo2010, Bellomo2010-2}. 

Partial knowledge of the global dynamics, inferred from the evolution of the system, can also be used to more efficiently predict future dynamics. This is embodied in the transfer tensor method~\cite{CerrilloCao2014}, where short time system dynamics can be used to construct a discretized memory kernel, in the form of the eponymous transfer tensors. These can then be used to propagate the system to later times, simulating the long-time dynamics with exponentially fewer resources than other methods when the exact propagation from an underlying model is computationally complex~\cite{Rosenbach2016,Kananenka2016}. Memory kernels are often used to formulate exact, continuous time open dynamics, most famously in the form of the Nakajima-Zwanzig master equation~\cite{breuerpetruccione}. However, unless the underlying dynamics is homogeneous in time, and the system is initially uncorrelated with its environment, there is no clear way to construct the memory kernel or transfer tensors operationally, either in experiment or in numerical simulations.

In this Article, we start by solving this problem definitively: First, in Sec.~\ref{sec:operational} we derive transfer tensors from first principles for any open quantum dynamics, in particular those with initial correlations and generators that have arbitrary periodic or transient time dependence. We then go on, in Sec.~\ref{sec:numerical} to demonstrate the power of the transfer tensor technique for simulating long-time dynamics, explicitly recovering the dynamical steady state of a driven, dissipative example system. In Sec.~\ref{sec:tteqnz}, we construct a master equation from the transfer tensors, demonstrating a direct correspondence between tomographically reconstructed dynamics and a generalised Nakajima-Zwanzig master equation. Finally, in Sec.~\ref{sec:operationalmeaning}, we show how this correspondence could be used in an experimental setting to relate directly measurable quantities to properties of the underlying dynamics. 

Our results render a large class of difficult numerical problems efficiently solvable and build a new link between underlying physics and experimentally accessible quantities. In addition, they open up the possibility of deriving approximate master equations for the system, through operationally meaningful coarse-graining of the reduced dynamics. Before presenting our main results, we first introduce memory kernels and their relation to system's equation of motion.

%%%%%%%%%%%%%%%%%%%%%%%%%%%%%%%%%%%%
%%%%%%%%%%%%%%%%%%%%%%%%%%%%%%%%%%%%
%%%%%%%%%%%%%%%%%%%%%%%%%%%%%%%%%%%%
%%%%%%%%%%%%%%%%%%%%%%%%%%%%%%%%%%%%

%%%%%%%%%%%%%%%%%%%%%%%%%%%%%%%%%%%%
%%%%%%%%%%%%%%%%%%%%%%%%%%%%%%%%%%%%
\subsection*{Memory kernel master equations}
\label{sec:memkern}
%%%%%%%%%%%%%%%%%%%%%%%%%%%%%%%%%%%%
%%%%%%%%%%%%%%%%%%%%%%%%%%%%%%%%%%%%

The most general form for the generator of a memoryless, or Markov, quantum evolution was famously derived by Lindblad~\cite{Lindbladeqn}, as well as Gorini, Kossakowski and Sudarshan~\cite{GKSeqn}. In this case, the derivative of the system state at time $t$ is a homogeneous linear function of the state at that same time, as in Eq.~\eqref{eq:underlying}. For more general, non-Markovian processes, one natural way to extend the formalism is to consider master equations in which the time derivative depends on the history of the system's evolution:
\begin{gather}
\dot{\rho}_t = \mathcal{L}_t \rho_t + \int_{t_0}^{t}\rmd s\, \mathcal{K}_{t,s} \rho_s + \mathcal{J}_{t,t_0}, \label{eq:memkernME}
\end{gather}
where $\mathcal{L}_t$ is a time-local generator, $\mathcal{K}_{t,s}$ is a superoperator known as the memory kernel, and $\mathcal{J}_{t,t_0}$ is an inhomogeneous term, that can depend on the initial condition. Put simply, the memory kernel $\mathcal{K}_{t,s}$ quantifies how the system state at time $s$ influences, through interaction with the environment, the evolution at later time $t$. Knowledge of the functions on the right hand side of Eq.~\eqref{eq:memkernME} allows for the self-consistent solution of the system's dynamics.

In recent years, several phenomenological master equations with memory kernels have been proposed~\cite{BarnettStenholm2001, DafferWodkiewicz2004, BreuerVacchini2008, ChruscinskiKossakowski2016, Vacchini2016}. It is always possible, given a microscopic description, to cast any open dynamics in the form of Eq.~\eqref{eq:memkernME} using the Nakajima-Zwanzig projection superoperator technique~\cite{breuerpetruccione} (see Sec.~\ref{sec:tteqnz} and Appendix~\ref{app:NZTDOP}), though the solution of the resulting equation is far from trivial and calculating the memory kernel for different models remains an ongoing research problem~\cite{Geva2003,Cohen2011,Cohen2013}. However, there is currently no prescription for determining the elements of Eq.~\eqref{eq:memkernME} when a microscopic model is unavailable. In Sec.~\ref{sec:tteqnz} we show how this can be done, in a simulation or an experiment, in terms of completely positive dynamical maps, but first we show how a discrete time memory kernel structure arises, for general open dynamics, in the form of transfer tensors.

%%%%%%%%%%%%%%%%%%%%%%%%%%%%%%%%%%%%
\section{Operational derivation of transfer tensors} 
\label{sec:operational}
%%%%%%%%%%%%%%%%%%%%%%%%%%%%%%%%%%%%

The transfer tensor method was first proposed as an ansatz in Ref.~\cite{CerrilloCao2014} under certain restrictive conditions: (i)~the underlying dynamics is time homogeneous, i.e., $\mathcal{L}^\se_t = \mathcal{L}^\se$; (ii)~the initial $\setxt$ state is uncorrelated: $\rho^\se_{t_0}=\rho_{t_0}\otimes \rho^\e_{t_0}$; (iii)~$\rho^\e_{t_0}$ is a stationary state. Our key result is a first principles derivation of the transfer tensors for a general open evolution, without making any of the above assumptions. 

We begin by introducing a set of objects $\{\mathbf{P}_s, \mathbf{Q}_s\}$ which act as
\begin{gather}\label{eq:opPQ}
\mathbf{P}_s\rho_{t} = \Lambda_{t:s} \rho_{s}
\quad \mbox{and} \quad 
\mathbf{Q}_s\rho_{t} = \rho_{t}-\Lambda_{t:s} \rho_{s}, 
\end{gather}
such that $(\mathbf{P}_s+\mathbf{Q}_s)\rho_{t}=\rho_{t}$. Subsequent action of these objects further breaks up the evolution, i.e., $\mathbf{P}_{s'} \Lambda_{t:s} =\Lambda_{t:s'}\Lambda_{s':s}$.  Here, $\Lambda_{t:s}$ is a completely positive dynamical map from time $s$ to time $t$, $\Lambda_{t:s}\rho = \tre \{\mathcal{U}^\se_{t:s} (\rho \otimes \tau^\e_s) \}$, where, as in the introduction, $\mathcal{U}^\se_{t:s}$ is the time evolution operator of the underlying dynamics and $\tau^\e_s$ is a unit trace time-dependent density operator not necessarily related to the actual environment dynamics; we will elaborate on the interpretation of $\Lambda_{t:s}$ and $\tau^\e_s$ in numerical and experimental contexts in Secs.~\ref{sec:numerical}~and~\ref{sec:operationalmeaning} respectively; for now, $\Lambda_{t:s}$ can be thought of simply as a two parameter family of dynamical maps. 

Notably, $\mathbf{P}_s$ and $\mathbf{Q}_s$ are not superoperators, rather they can be seen as similar kinds of objects to the time ordering operator; that is, as consistently defined prescriptions for rewriting the subsequent expression based on its time indices. $\mathbf{P}_s$ replaces an object at any $t>s$  (either a dynamical map from $s'<s$ to $t$ or the system state at time $t$) with that at time $s$ acted on by $\Lambda_{t,s}$; $\mathbf{Q}_s$ gives the complement. These can be seen simply as a shorthand for the decomposition $\rho_t = \Lambda_{t:s}\rho_s +(\rho_t - \Lambda_{t:s}\rho_s)$ (and a corresponding decomposition for maps), which we will now iterate to further break up the dynamics. 
Specifically, by repeatedly using the identity $(\mathbf{P}_s + \mathbf{Q}_s) = \mathcal{I}$, we can write the evolution up to time $t=t_N$ in terms of the evolution to $N-1$ earlier times $\{t_j\}$
\begin{widetext}
\begin{align}
\label{eq:superchanneldecomposition}
\rho_{t}=& \left(\mathbf{P}_{t_{N-1}} + \mathbf{Q}_{t_{N-1}}\left(\mathbf{P}_{t_{N-2}} + \mathbf{Q}_{t_{N-2}} \right.\right. 
\left(\dots
\left(\mathbf{P}_{t_{1}} + \mathbf{Q}_{t_{1}}
\left(\mathbf{P}_{t_{0}} + \mathbf{Q}_{t_{0}} \right)\dots\right)\right)
\rho_t\\
%%%%%%%%%%%%%%%%%%
=& \sum_{j=0}^{N-1} T^{(N-j)}_{t:t_j}\rho_{t_j} + \Xi_{t:t_0}, \label{eq:Mdecomp}
\; \mbox{where} \; \,
T^{(N-j)}_{t:t_j}\rho_{t_j} =  \mathbf{Q}_{t_{N-1}} \cdots \mathbf{Q}_{t_{j+1}} \mathbf{P}_{t_{j}}\rho_t 
\; \mbox{and} \;\,
\Xi_{t:t_0} = \mathbf{Q}_{t_{N-1}} \cdots \mathbf{Q}_{t_{0}} \rho_t.
\end{align}
\end{widetext}

Here, the $T^{(n)}_{t,s}$ are exactly the transfer tensors defined in Ref.~\cite{CerrilloCao2014}, except that we are able to explicitly derive them in terms of, in principle, operationally accessible maps, without appealing to time-homogeneity of the memory kernel. They are given by
\begin{align}
& T^{(1)}_{t:t_{N-1}} = \Lambda_{t:t_{N-1}} \quad \mbox{and} \nonumber \\
& T^{(N-j)}_{t:t_j} = \Lambda_{t:t_j}  - \sum_{k=j+1}^N T^{(N-k)}_{t:t_k} \Lambda_{t_k:t_j}.
\label{eq:transfertensor}
\end{align}
Chasing this definition leads to the explicit form 
\begin{align}
T^{(N-j)}_{t:t_j}\!=& \Lambda_{t:t_j} -\sum_{k=j+1}^{N-1} \Lambda_{t:t_k}\Lambda_{t_k:t_j} \nonumber \\
&+\!\sum_{k=j+2}^{N-1} \sum_{l=j+1}^{k-1} \Lambda_{t:t_k}\Lambda_{t_k:t_l}\Lambda_{t_l:t_j}\! -\dots. 
\end{align}

Finally, the remaining inhomogeneous term in Eq.~\eqref{eq:Mdecomp} encodes all memory effects arising from the initially correlated $\setxt$ state, and is defined as the difference $\Xi_{t:t_0} = \rho_t - \sum_{j=0}^{N-1} T^{(N-j)}_{t:t_j}\rho_{t_j}$; this can be operationally determined for any initial preparation by reconstructing the superchannel $\mathcal{M}_{t:t_0}$. For long enough times, this term can be neglected~\cite{BuserCerrillo2017}; however, in Appendix~\ref{app:initialcorr} we show how this term can be included without direct calculation at the expense of reconstructing a larger set of dynamical maps. There, we account for the initial correlations operationally by replacing the map $\Lambda_{t:t_0}$ with the completely positive superchannel $\mathcal{M}_{t:t_0}$~\cite{Modi2012}. We will now show explicitly how the transfer tensors derived in this section can be used to simulate driven open systems.

%%%%%%%%%%%%%%%%%%%%%%%%%%%%%%%%%%%%
%%%%%%%%%%%%%%%%%%%%%%%%%%%%%%%%%%%%
\section{Efficiently simulating long-time dynamics}
\label{sec:numerical}
%%%%%%%%%%%%%%%%%%%%%%%%%%%%%%%%%%%%
%%%%%%%%%%%%%%%%%%%%%%%%%%%%%%%%%%%%

The principle advantage of the transfer tensor approach over other simulation methods is that it allows for the determination of long-time dynamics, with an error that does not grow with the evolution time~\footnote{Assuming that the short time exact dynamics is reconstructed accurately.} but instead depends on the time beyond which memory effects are neglected. This is in contrast to many numerically exact techniques, where the computational effort to achieve a given numerical precision grows exponentially with the evolution time~\cite{devega2017} (of course, if memory effects persist beyond the time taken to converge on a steady state, then the numerically exact techniques will do just as well). In what follows, we will outline how a transfer tensor simulation could be performed and how the error can be bounded.

Fixing $\delta t = t_{j+1}-t_j$ determines the time resolution of the resulting dynamics, this is an external choice that does not affect the accuracy of the approach, but rather limits the degree to which fine-grained dynamical features can be resolved. To guarantee all such features are captured, one could choose it to be smaller than the time scale set by the largest energy in the system to be simulated. Once the temporal resolution is set, we will assume that the dynamical maps $\Lambda_{t:s}$ are periodic (or transient), such that $\Lambda_{t+T:s+T}=\Lambda_{t:s}$ for large enough $s$, and where the period $T=c \, \delta t$ with integer $c$. This is equivalent to requiring that the underlying generator $\mathcal{L}^\se_t$ and reference state $\tau^\e_t$ are transient or periodic with periods $a \, \delta t$ and $b \, \delta t$, where $c$ is the greatest common denominator of $1/a$ and $1/b$ ($\delta t$ and $\tau^\e_t$ can always be chosen in a simulation to be compatible with the generator in this way). In this case, more general than that considered in Ref.~\cite{CerrilloCao2014}, we have periodicity of the transfer tensors, such that 
\begin{gather}
T^{(l)}_{t_{k}:t_{k-l}} = T^{(l)}_{t_{k+c}:t_{k-l+c}}, \quad \forall \, l<k \;\, \mbox{and} \;\, \forall \, t_{k-l}
\end{gather}
sufficiently large that transient effects are negligible. Therefore, only transfer tensors $T^{(l)}_{t_{k}:t_{k-l}}$ with $k-l<c$ need to be calculated explicitly (in the transient case, all transfer tensors starting within the transient period must be computed).

Furthermore, in most real open systems, the memory kernel $\mathcal{K}_{t,s}$ decreases in magnitude (often exponentially) with the time difference $t-s$~\cite{breuerpetruccione}. That is, the system's state at one time is continuously forgotten until it no longer affects the future evolution. Consequently, the transfer tensors $T^{(l)}_{t_{k}:t_{k-l}}$ also become negligibly small for sufficiently large $l$. Similarly, the influence of the initial state encoded in $\mathcal{J}_{t,t_0}$ or $\Xi_{t,t_0}$ commonly decays with $t-t_0$. What this means is that, beyond a certain time $t_m=m \, \delta t$, memory effects can be neglected. That is, $T^{(l)}_{t_{k}:t_{k-l}}$ with $l>m$ can be effectively set to zero without substantially affecting the dynamics. Combined with the time translation symmetry described above, this means the dynamics at any time, given the finite resolution set by $\delta t$, can be reconstructed using only a finite set of $cm$ transfer tensors ($m$ transfer tensors for each of the $c$ starting points in a driving period). As such, decreasing $\delta t$ while leaving other time scales fixed increases the computational cost quadratically.

In the first step of the simulation, numerical process tomography is performed by exactly solving the dynamics to a time $t_m$ beyond each time $x \delta t$, $x<c$, in the first driving period (or the transient period) for an informationally complete set of initial states. These dynamics can be reconstructed for any choice of environment state $\tau^\e_t$ permitted by the exact technique used to solve for the short time dynamics (though different choices can lead to longer or shorter memory times, as discussed in Sec.~\ref{sec:operationalmeaning}). The resulting set of dynamical maps is then used to reconstruct the transfer tensors on this interval (starting with those between adjacent time steps and iterating with Eq.~\eqref{eq:transfertensor}). Importantly, despite the fact the transfer tensors are constructed starting from different points of the driving cycle, the exact dynamics only needs to be calculated for an evolution time $t_m$ from each point, meaning the efficiency of the technique does not depend on the driving period.  Using Eq.~\eqref{eq:Mdecomp}, the state at any arbitrary time $t_k$ can be found by identifying $T^{(l)}_{t_{k}:t_{k-l}}$ with $T^{(l)}_{t_{k-l}\Mod T+l\delta t:t_{k-l}\Mod T}$, for $l\leq m$, and with $0$, for $l>m$ (similarly, $\Xi_{t_k,t_0}$ is set to $0$ for $k>m$), before propagating the initial state. We reiterate that the complexity of this latter stage is independent of the underlying $\setxt$ model, meaning it could be equally well applied to any open system whose short time dynamics can be determined.

In Appendix~\ref{app:error}, we show how the maximum distance between the numerically propagated state $\tilde{\rho}_{t_k}$ and the exact state $\rho_{t_k}$ can be bounded as 
\begin{align}
&\left\|\rho_{t_k}\!-\tilde{\rho}^{(m)}_{t_k}\right\|_1 \nonumber\\
&\quad\!\lesssim\! \sum_{l=1}^{m}\left\|T^{(2m-l)}_{t_{k-2m} \Mod{T}+2t_m:t_{k-2m} \Mod{T}+l\delta t}\right\|, \label{eq:errorbound}
\end{align}
where $\|X\|:= \max_{\rho,\sigma}\tr\{\sigma X \rho\}$ is the operator norm of the matrix representation of superoperator $X$ that acts on vectorised density operators. In other words, the error grows with the cumulative size of the transfer tensors beyond the first memory time.  What this means is that one can guarantee a desired accuracy in the propagated state by varying $t_m$ and constructing transfer tensors to time $T+2t_m$ until the right hand side of the above equation is small enough.
The upper bound is approximate, insofar as it only considers contributions to the error from the second memory time period (as long as the memory decays faster than arithmetically, then contributions from subsequent memory times will not diverge). However, we find that this bound tends to grossly overestimate the propagation error, which is more closely bounded by the largest norm of the `longest' transfer tensors: ${\rm max}_{x\leq c}\|T^{(m)}_{x\delta t+t_m:x\delta t}\|$.  On the other hand, if the memory cutoff approximation is too severe, then the propagated dynamics may become unphysical, leading to divergent error in some cases (as in other approximate methods, where throwing away significant terms results in unphysical behaviour~\cite{breuerpetruccione, Fruchtman2016}). 

%%%%%%%%%%%%%%%%%%%%%%%%%%%%%%%%%%%%
%%%%%%%%%%%%%%%%%%%%%%%%%%%%%%%%%%%%
\begin{figure}[thp]
\centering
\includegraphics[width=0.42\textwidth]{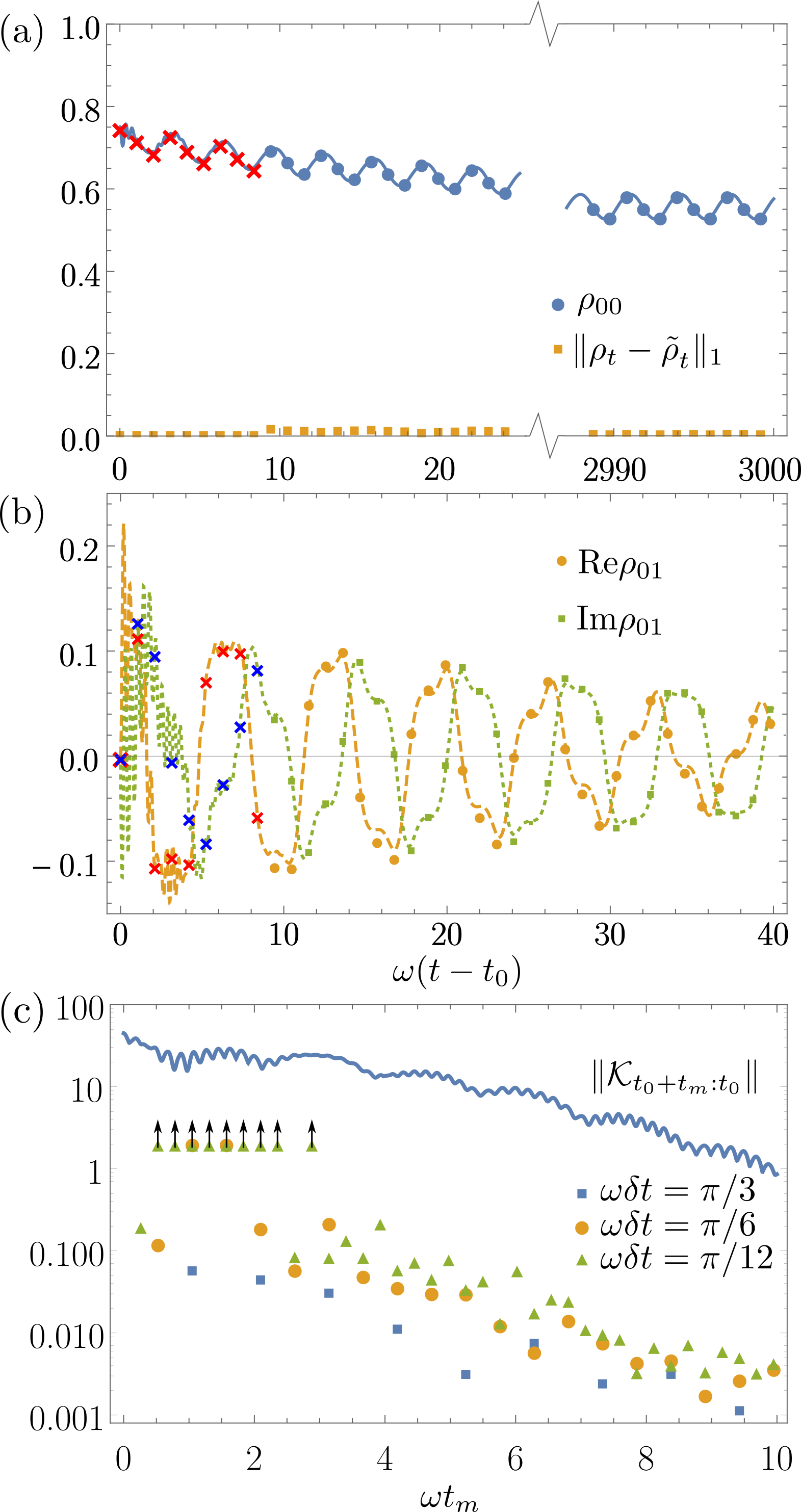}
\caption{\textbf{Example propagated dynamics} (a) elements of the system density operator at short and long times for model described in the text. Transfer tensors are reconstructed, using a time independent reference state, between the first nine time steps (red and blue crosses), before being used to propagate the state to later times. The propagated population of the system's excited state (blue dots) is shown alongside exactly calculated dynamics (continuous lines). Also shown (yellow squares) is the trace distance between the exact density operator and its transfer tensor propagated counterpart. (b) Real and imaginary parts of the system's coherence (the coherence is zero at long times). (c) Long time propagation error $\|\rho_t-\tilde{\rho}_t\|_1$ as a function of memory time $t_m$ and time step size $\delta t$; also plotted, for reference, is the norm of the memory kernel at $t_m$. Points with arrows correspond to cases where the distance is greater than two (sometimes by many orders of magnitude), indicating non-physical dynamics. In all cases, the bound in Eq.~\eqref{eq:errorbound} is larger than the error.}
\label{fig:dynamics}
\end{figure}
%%%%%%%%%%%%%%%%%%%%%%%%%%%%%%%%%%%%
%%%%%%%%%%%%%%%%%%%%%%%%%%%%%%%%%%%%

To illustrate the effectiveness of the transfer tensor we now consider a simple example of a driven, dissipative open system. The dynamics of the total system-environment state is governed by a memory-less master equation:
\begin{align}
\mathcal{L}^\se_t\rho^\se_t =& -i [H^\se_t,\rho^\se_t] \nonumber \\
&+ \Gamma\left(L\rho^\se_t L^\dagger-\frac{1}{2}\{L^\dg L,\rho^{\se}_t\}\right).
\end{align}
While the $\setxt$ dynamics is Markovian, the dynamics of $\stxt$ alone will be non-Markovian and will have the form of Eq.~\eqref{eq:memkernME}.

For concreteness, we take the system-environment to be a pair of qubits. The two evolve according to the Hamiltonian $H^\se_t = \frac{\omega}{2}\sigma_z^\s + \frac{\omega'}{2}\sigma_z^\e +g[\sigma_x^\s \otimes \sigma_x^\e+\cos(\Omega t)\sigma_y^\s \otimes \sigma_y^\e]$, and the Lindblad operators $L = \mathbbm{1}^\s\otimes \ketbra{0}{1}^\e$ act on the environment alone. This generates dissipative dynamics for the environment; incoherently pumping it to the excited state with rate $\Gamma$. This seemingly simple model has many of the features of more complex open systems; in fact, it is approximately equivalent to a class of spin-boson models~\cite{Tamascelli2018}.

The simulated dynamics, with correlated initial state $\rho^{\se}_0 = \frac{3}{4}\ketbra{0}{0}\otimes\ketbra{+}{+}+\frac{1}{4}\ketbra{1}{1}\otimes\ketbra{-}{-}$, where $\ket{\pm}=(\ket{0}\pm\ket{1})/\sqrt{2}$, is shown in Fig.~\ref{fig:dynamics} along with the accompanying error, quantified by the trace distance $\|\rho_{t_k}-\tilde{\rho}_{t_k}\|_1 = \tr\{|\rho_{t_k}-\tilde{\rho}_{t_k}|\}$ (in all figures the parameters are set as $\omega=\Gamma=\omega'/16=g/2=\Omega/2$). 

In the figure, short time exact dynamics, calculated between the first nine time steps (using a time-independent reference state $\tau^\e=\tr_\s\rho_0^\se$), is used to propagate the state $\tilde{\rho}_{t_k}$ to much longer times with no noticeable increase in error, as indicated by the yellow dots, which remain close to zero for all times. This can also be observed by comparing the exact dynamics (solid curves) with the propagated points. The dynamics are accurate despite choosing a relatively short memory cutoff time ($\omega t_m=5$). As shown in panel (b), the memory kernel still has a significant norm at this time, which would usually lead one to conclude that we are discarding important memory contributions when making the cutoff.

One of the reasons for the remarkable effectiveness of the transfer tensor approach, in this case, is the coarse sampling (long $\delta t$) of the initial dynamics. In general, as seen in panel (b) of the figure, where we calculate the long-time propagation error for a variety of cutoff times and time step lengths, we find that finer grained dynamics require longer memory times for the same accuracy \footnote{We could have also reduced the error by choosing a time-dependent reference state (since the memory kernel norm decays much faster). However, in this case, a longer memory time would need to be chosen to allow transients in the time-dependent state to decay.}. In every case, we find that the bound in Eq.~\eqref{eq:errorbound} is extremely loose, meaning that the transfer tensor propagation is more robust than indicated by the conservative analysis presented in Appendix~\ref{app:error}. However, beyond a certain point, even this loose bound on the error seems to decrease exponentially with $t_m$. In the limit of continuous time dynamics, our approach leads us to a generalised version of the Nakajima-Zwanzig equation, of the form of Eq.~\eqref{eq:memkernME}, as we will now show.

%%%%%%%%%%%%%%%%%%%%%%%%%%%%%%%%%%%
\section{Equivalence with the Nakajima-Zwanzig equation in the continuum limit}
\label{sec:tteqnz}
%%%%%%%%%%%%%%%%%%%%%%%%%%%%%%%%%%%

While Eq.~\eqref{eq:memkernME} governs the exact dynamics of the system alone, conventional derivations usually rely implicitly on an underlying model for the $\setxt$ dynamics. In this sense, the evolution it predicts is only as good as the initial model, which may only approximate the system's true behaviour. We will show how operationally determined quantities, encoded in transfer tensors, can be used to construct discrete-time memory kernel equations for general open processes, where the $\setxt$ model may not be well known, and show that these tend to a generalised Nakajima-Zwanzig equation in the continuum limit.

To do this, we will construct a difference equation for the state at time $t$, before taking a limit and recovering a derivative at that time. We first divide the interval $[t_0,t]$ into $N$ parts, as above, and choose the $N+1$ times $\{t_k\}$ to each be $\delta t$ apart: $t_{j+1} - t_j =\delta t := (t-t_0)/N$ \footnote{Equidistance between time steps is not necessary for our derivation and is only chosen for convenience. It may be that other choices lead to faster convergence in practice.}. By substituting  Eq.~\eqref{eq:Mdecomp} into the difference between $\rho_t=\rho_{t_N}$ and $\rho_{t_{N-1}}$ we get:
\begin{widetext}
\begin{align}
\frac{\rho_{t}-\rho_{t_{N-1}}}{\delta t} =  \frac{(\Lambda_{t:t_{N-1}}-\mathcal{I})}{\delta t}\rho_{t_{N-1}}
+\sum_{j=0}^{N-2}\delta t\, \frac{T^{(N-j)}_{t:t_j}}{\delta t^2}\rho_{t_j}
 +\frac{\Xi_{t:t_0}}{\delta t}, \label{eq:operationalderivative}
\end{align}
\end{widetext}
which resembles, term by term, a discrete version of the memory kernel master equation given in Eq.~\eqref{eq:memkernME}. In the limit $N \to \infty$ we can equate the terms as:
\begin{align}
&\mathcal{K}_{t,t_j} = \lim_{N\rightarrow\infty} \frac{T^{(N-j)}_{t:t_j}}{\delta t^2 }, \quad \mathcal{J}_{t,t_0} = \lim_{N\rightarrow\infty}\frac{\Xi_{t:t_0}}{\delta t}, \nonumber \\
& \mathcal{L}_{t} = \lim_{N\rightarrow\infty} \frac{\Lambda_{t:t_{N-1}}-\mathcal{I}}{\delta t}. \label{eq:opkernel} 
\end{align}

In order to relate these quantities directly to the underlying dynamics, in terms of $\mathcal{L}^\se_t$ and $\rho_0^\se$, we can substitute $\Lambda_{t:s}\rho = \tre \{\mathcal{U}^\se_{t:s} (\rho \otimes \tau^\e_s) \}$ into the left hand sides of these expressions (writing $T^{(N-j)}_{t:t_j}$ and $\Xi_{t:t_0}$ in terms of dynamical maps). In this way, we can recast the objects in the decomposition in Eq.~\eqref{eq:superchanneldecomposition} in terms of $\setxt$ projection superoperators as 
\begin{align}
& \mathbf{P}_s\rho_t =\tre\{\mathcal{U}^\se_{t:s} \mathcal{P}^\se_s \mathcal{U}^\se_{t:t_0} \mathcal{A} \rho^\se_{t_0}\}
\quad \mbox{and} \nonumber \\
& \mathbf{Q}_s\rho_t = \tre\{\mathcal{U}^\se_{t:s} \mathcal{Q}^\se_s \mathcal{U}^\se_{t:t_0} \mathcal{A}\rho^\se_{t_0}\},
\end{align}
where $\mathcal{P}^\se_s\rho^\se_{s} = \tre\{\rho^\se_{s}\}\otimes \tau_s^\e$ and $\mathcal{Q}^\se_s=\mathcal{I}^\se-\mathcal{P}^\se_s$, with $\tau_s^\e$ the environment state appearing in the underlying representation of the dynamical map $\Lambda_{t,s}$; $\mathcal{A}$ is, as before, the operation used to prepare the system at the initial time. We can then take the limit by expanding $\mathcal{U}^\se$ for small $\delta t$: $\mathcal{U}^\se_{t+\delta t:t} \simeq \mathcal{I}^\se + \delta t \mathcal{L}^\se_t + \delta t^2 {\mathcal{L}^\se_t}^2+\dots$. We do this in detail in Appendix~\ref{app:expTT}, finding 
\begin{align}
&\mathcal{L}_t\rho_t=\tre\left\{\mathcal{P}^\se_t \mathcal{L}^\se_t\mathcal{P}^\se_t\rho_{t}\otimes x^\e\right\} \nonumber\\
&\mathcal{K}_{t,s}\rho_{s} = \tre\left\{\mathcal{P}^\se_t \mathcal{L}^\se_t \mathcal{G}_{t,s} 
\phantom{\left(\dot{\mathcal{Q}}^\se_{s} \right)}
\right. \label{eq:memkern} \\
& \nonumber \phantom{x}\qquad\qquad \times  \left.\left(\mathcal{Q}^\se_{s} \mathcal{L}^\se_{s} \mathcal{P}^\se_{s}-\dot{\mathcal{P}}^\se_{s} \right) 
\rho_{s}\otimes x^\e\right\},\\
& \mathcal{J}_{t,t_0} = \tre\left\{\mathcal{P}^\se_t \mathcal{L}^\se_t \mathcal{G}_{t,t_0}\mathcal{Q}^\se_{t_0}\mathcal{A}\rho^\se_{t_0}\right\} %\label{eq:inhomterm}
\nonumber,
\end{align}
where the propagator $\mathcal{G}_{t,s}=T_{\leftarrow} \exp\left[\int_{s}^t\rm{d} s'\, \mathcal{Q}^\se_{s'}\mathcal{L}^\se_{s'}\right]$ is a time ordered exponential (as indicated by $T_{\leftarrow}$) and the value of the unit trace operator $x^\e$ is unimportant, as it is acted on directly by a projector.

Equation~\eqref{eq:memkern} represents exactly the terms appearing in a generalised version of the Nakajima-Zwanzig master equation. In the standard derivation of the Nakajima-Zwanzig equation, a time-independent superoperator $\mathcal{P}^\se$ and its complement $\mathcal{Q}^\se=\mathcal{I}^\se-\mathcal{P}^\se$ are defined on the joint $\setxt$ space~\cite{breuerpetruccione}. In Appendix~\ref{app:NZTDOP}, we show that one arrives at Eq.~\eqref{eq:memkern} following the standard derivation, but starting with the time-dependent projection superoperators defined above.
Unlike in earlier approaches utilising time-dependent projectors for open systems (see e.g. Refs.~\cite{WillisPicard1974,PicardWillis1977}) we do not place any restrictions on the operator $\tau_t^\e$ beyond that it should be a positive, unit trace density operator, and that it and its first derivative $\dot{\tau}_t^\e$ should be continuous functions of time. It is worth noting the similarity to recent projection operator approaches to describing non-linear many-body dynamics~\cite{DegenfeldSchonburg2014,DegenfeldSchonburg2015}.

The significance of our result is that, even for time-inhomogeneous $\setxt$ dynamics and initial correlations, the memory kernel can be operationally determined by reconstructing the short time-difference dynamics starting from a series of intermediate time points, alleviating the need to explicitly calculate or approximate the time-ordered exponential integrals implicit in~Eq.~\eqref{eq:memkern}. However, the convergence of the limit in Eqs.~\eqref{eq:opkernel} may be slow, and will depend on the time difference $t-t_j$ in general.

\begin{figure}[ht]
\centering
\includegraphics[width=0.48\textwidth]{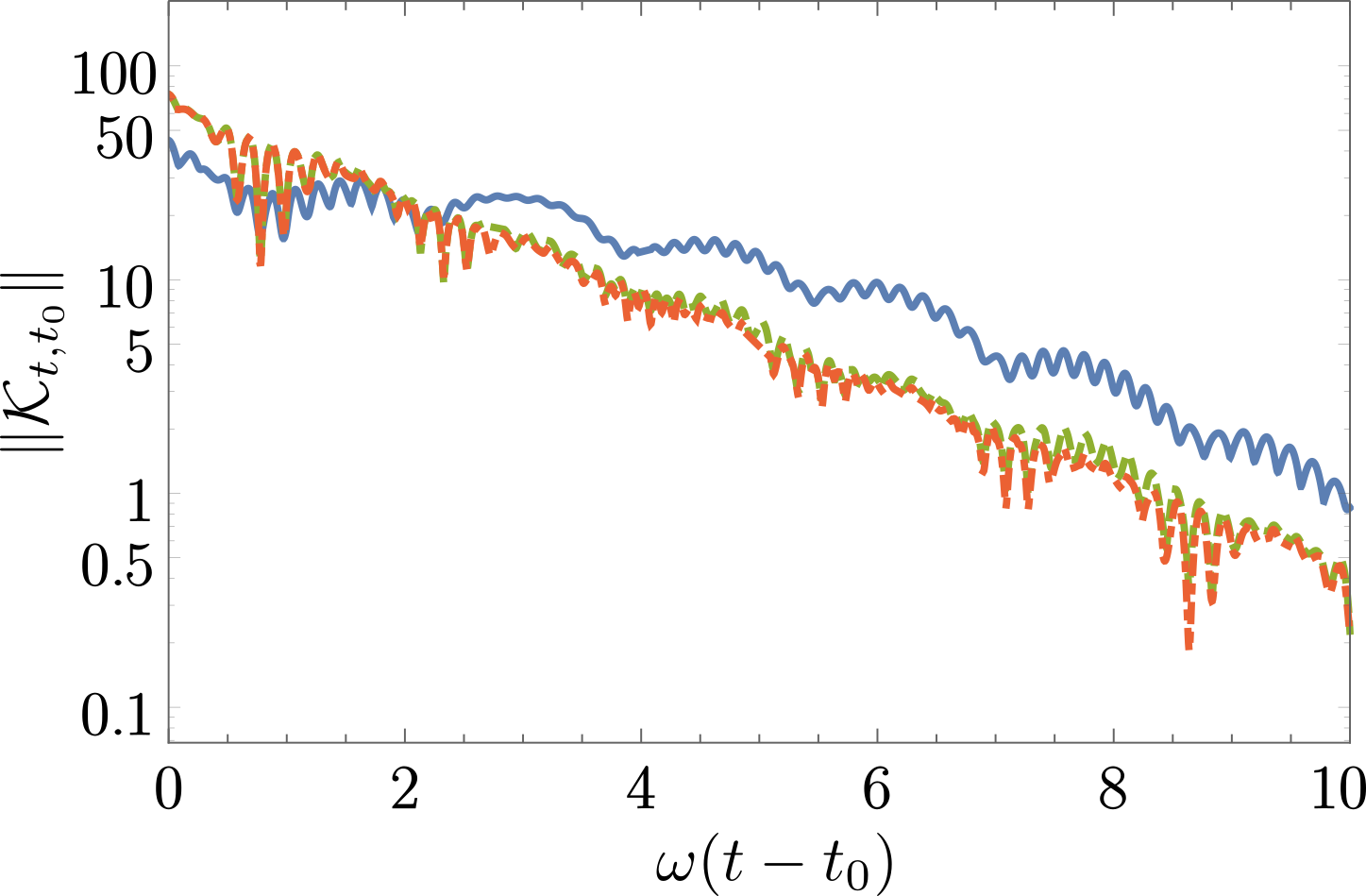}
\caption{\textbf{Norm of different memory kernels for the same process.} Derived from exact simulation of the model presented in Sec.~\ref{sec:numerical}. Three different choices of $\mathcal{P}^\se_t$: (i) the usual case of projection onto initial environment state, $\mathcal{P}^\se \rho^\se_t=\rho_t\otimes \trs\{\rho^\se_{t_0}\}$ (blue, solid, largest at long times); (ii) projection arising from keeping the system in fixed state $\ketbra{0}{0}$, i.e., $\mathcal{E}_t=\mathcal{B}_{\ketbra{0}{0}}$ (red, dot-dashed, smallest at long times); (iii) projection onto true environment state, $\mathcal{P}^\se_t\rho^\se_t=\rho_t\otimes \trs\{\rho^\se_{t}\}$, corresponding to $\mathcal{E}_t=\mathcal{I}$ (green, dashed). Note the log scale.}
\label{fig:memkernexample}
\end{figure}

Importantly, there is a freedom in choosing the environment operator $\tau_t^\e$ that goes into the definition of the dynamical map $\Lambda_{t,s}$ and, hence, the projector $\mathcal{P}^\se_t$. Different choices of projector $\mathcal{P}^\se_t$ lead to different forms for the superoperators in Eqs.~\eqref{eq:memkern} but the same solution for the dynamics $\rho_t$. This is exemplified, using the model presented in Sec.~\ref{sec:numerical}, in Fig.~\ref{fig:memkernexample}, where different (time-dependent) projectors are seen to lead to memory kernels with different norms; in this case, a particular time-dependent choice leads to a memory kernel whose norm decays several times faster than the most obvious time-independent choice (where $\tau_t^\e=\tr_\s\rho_0^\se$). For simulations, this means that a judicious choice of projector can lead to a much shorter memory cutoff time for the same degree of approximation in the dynamics. 
While $\tau_t^\e$ can be freely chosen in a simulation (exact numerical method permitting), the dynamical maps reconstructed in an experimental setting have an environment operator dictated by the $\setxt$ dynamics. However, as we will now discuss, there is still an analogous freedom in how the tomography is performed.

%%%%%%%%%%%%%%%%%%%%%%%%%%%%%%%%%%%%
%%%%%%%%%%%%%%%%%%%%%%%%%%%%%%%%%%%%
%%%%%%%%%%%%%%%%%%%%%%%%%%%%%%%%%%%%
%%%%%%%%%%%%%%%%%%%%%%%%%%%%%%%%%%%%

%%%%%%%%%%%%%%%%%%%%%%%%%%%%%%%%%%%
\section{Experimental interpretation of projector choice} \label{sec:operationalmeaning}
%%%%%%%%%%%%%%%%%%%%%%%%%%%%%%%%%%%

One of the applications of the operational derivation of the Nakajima-Zwanzig equation, presented in the previous subsection, is to relate experimentally accessible quantities to properties of the underlying $\setxt$ Hamiltonian and initial state (through Eqs.~\eqref{eq:opkernel}~\&~\eqref{eq:memkern}). In this case, it is important to identify exactly which choices of environment state $\tau^\e_t$ and hence dynamical map $\Lambda_{t,s}$ (or equivalently projector $\mathcal{P}^\se_t$) correspond to which experimental reconstruction procedure, if any.

In the simplest scenario, the joint $\setxt$ evolution is left unperturbed before tomographically reconstructing $\Lambda_{t:s}$ (by discarding the state at time $s$ and preparing a fresh one, before measuring the system at time $t$). The state of the environment at time $s$, in this case, would be the freely evolved reduced state: $\rho^\e_s= \trs \{ \mathcal{U}^\se_{s:t_0} \rho^\se_{t_0}\}$. While this is perfectly reasonable from an operational standpoint, it is somewhat limiting from the perspective of extracting information about $\mathcal{L}^\se_t$ from the memory kernel. $\tau^\e_t$ must be known to separate out properties of the generator from the memory kernel, and the time-dependent state of the environment is (usually) at least as difficult to determine as $\rho_t$.

\begin{figure}[thp]
\centering
\includegraphics[width=0.46\textwidth]
{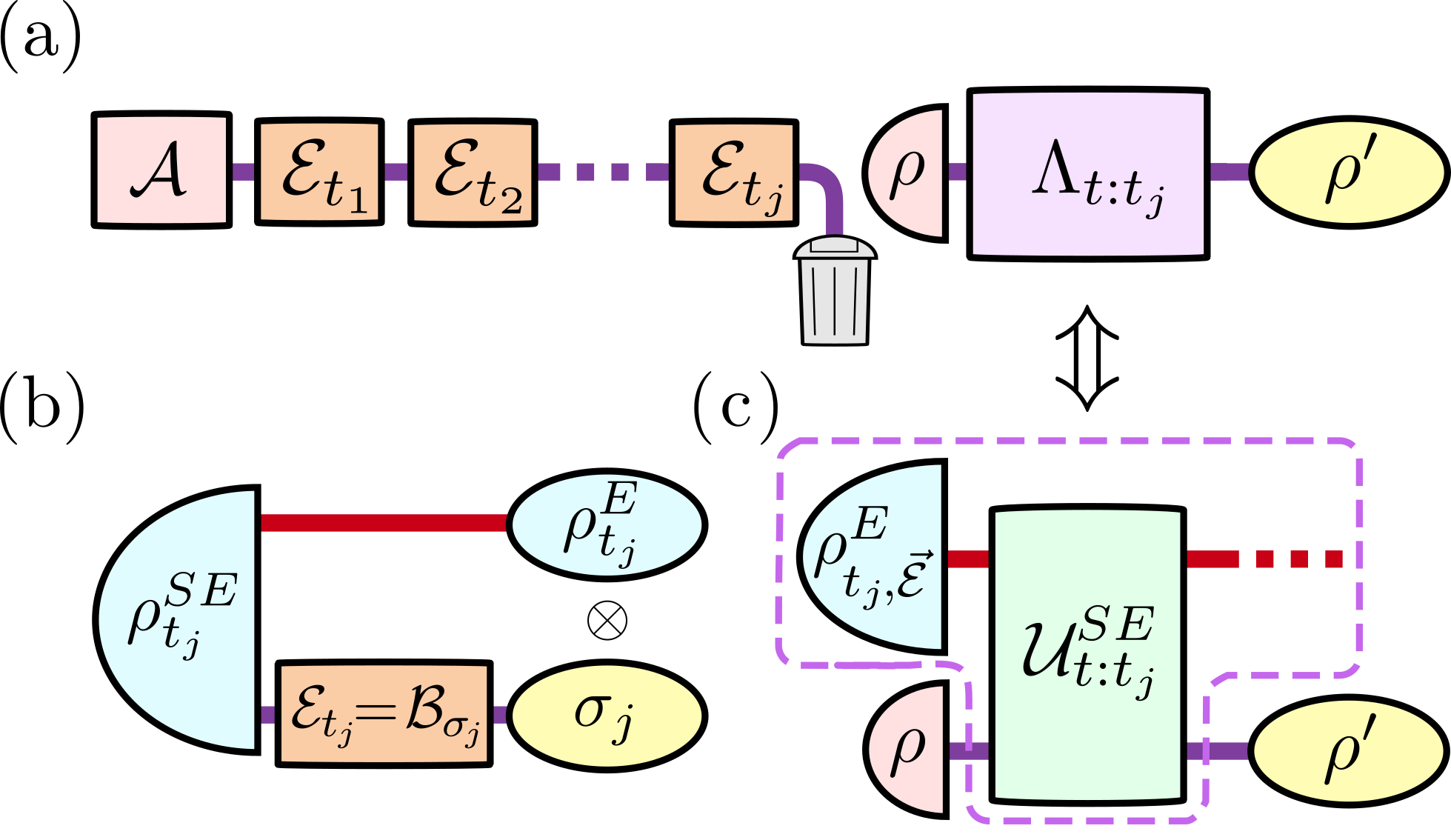}
\caption{\textbf{Experimentally reconstructing intermediate dynamics.} (a) After decorrelating the system from its environment (e.g., by projectively measuring and forgetting the outcome) at some intermediate time $t_j$, the subsequent dynamics is described by a CPTP map $\Lambda_{t:t_j}$. This can be operationally determined through normal quantum process tomography, i.e., by repreparing the system in various states $\rho$ and determining the corresponding $\rho'$ at later times. In general, operations $\vec{\mathcal{E}}=\{\mathcal{E}_{t_k}\}$ could be performed on the system at a series of earlier times $\{t_k<t_j\}$ (in addition to the initial preparation operation $\mathcal{A}$), leading to different $\Lambda_{t:t_j}$. (b) One choice of $\vec{\mathcal{E}}$ which leads to a self-consistently solvable reconstructed master equation is $\mathcal{E}_{t_k}=\mathcal{B}_{\sigma_k}$, where $\mathcal{B}_{\sigma_k}\rho=\sigma_k$. (c) The map $\Lambda_{t:t_j}$ corresponds to joint evolution with an environment initially in the history dependent state $\rho^E_{t_j,\vec{\mathcal{E}}}$.}
\label{fig:tomography}
\end{figure}

In order to get around this problem, one could choose to act on the system with a series of superoperators $\vec{\mathcal{E}}=\{\mathcal{E}_{t_k}\}$ (any physically allowed transformation of the system) between times $t_0$ and $s$, leading to different environment states $\rho^\e_{s,\vec{\mathcal{E}}}$ and, correspondingly, different families of intermediate dynamical maps $\Lambda_{t:s}$ (where we have left the $\vec{\mathcal{E}}$ dependence implicit for notational convenience). This procedure is depicted in Fig.~\ref{fig:tomography} and represents the operational analogue of the freedom to choose $\mathcal{P}^\se_t$ in Eqs.~\eqref{eq:memkern}. 

%%%%%%%%%%%%%%%%%%%%%%%%%%%%%%%%%%%%
%%%%%%%%%%%%%%%%%%%%%%%%%%%%%%%%%%%%
%%%%%%%%%%%%%%%%%%%%%%%%%%%%%%%%%%%%
%%%%%%%%%%%%%%%%%%%%%%%%%%%%%%%%%%%%

A particularly convenient choice (from the perspective of calculating $\rho^E_{t_j,\vec{\mathcal{E}}}$) is the one where entanglement-breaking operations $\mathcal{E}_{t_j}\rho=\mathcal{B}_{\sigma_{t_j}}\rho = \sigma_{t_j}$ are applied at times $t_j<s$, where $\sigma_{t_j}$ is some fixed state for each time step. In this case, the evolution of $\rho^\e_{t_j,\vec{\mathcal{E}}}$ effectively decouples from the system in the limit that these operations are applied infinitely quickly. That is, the reduced state of the environment evolves according to an average generator: $\dot{\rho}^\e_t = \trs\left\{\mathcal{L}^\se_{t} \sigma_t\otimes \rho^\e_t \right\} = \mathcal{L}^\e_t \rho^\e_t$. In an experiment, this could be achieved by, for instance, dynamically decoupling the system~\cite{Viola1999} or using the quantum Zeno effect to freeze the dynamics through frequent strong measurement~\cite{Facchi2008}, though these may be difficult in practice and represent only two possible choices of control that could lead to a known $\rho^E_{t_j,\vec{\mathcal{E}}}$. In many cases, there will be choices of $\sigma_t$ such that $\rho^\e_{t_0}$ is a stationary state with respect to $\mathcal{L}^\e_t$, leading to time-independent projectors $\mathcal{P}^\se$ and $\mathcal{Q}^\se$.

%%%%%%%%%%%%%%%%%%%%%%%%%%%%%%%%%%%%
%%%%%%%%%%%%%%%%%%%%%%%%%%%%%%%%%%%%
\section{Discussion}
%%%%%%%%%%%%%%%%%%%%%%%%%%%%%%%%%%%%
%%%%%%%%%%%%%%%%%%%%%%%%%%%%%%%%%%%%

In this Article, we have presented a derivation of the transfer tensor method, leading to a scheme for relating an operationally meaningful description of any open quantum process, in terms of completely positive dynamical maps, to a Nakajima-Zwanzig equation that depends on the underlying system-environment dynamics. We have also generalised the latter from the usual case to include time-dependent projectors. In addition to providing a fundamental connection between two different pictures of open dynamics, our result opens up the possibility for efficient simulation of the long-time evolution of driven systems or those with initial correlations. 

Instead of directly solving the dynamics of a system to some long time $t$, dynamical maps can be reconstructed to a point where the memory kernel has decayed (from the end of the first driving period). These can then be used to calculate transfer tensors which will propagate the system to long times with an error that depends only on the accuracy of the initial reconstruction and the memory cutoff approximation used. This could be used to, for instance, determine dynamical steady states of driven systems~\cite{Stace2013} or probe signatures of many-body localisation~\cite{MBL}.

Using Eq.~\eqref{eq:operationalderivative}, an experimentally or numerically reconstructed set of transfer tensors could also be used to create an approximate memory kernel through interpolation or form fitting. The resulting master equation would be guaranteed to give physically sensible solutions. Since its accuracy would depend on the smoothness of the memory kernel rather than, e.g., the strength of the $\setxt$ coupling, it would provide an alternative to the usual perturbative approaches to approximate dynamics.

Finally, since the memory kernel contains information about the $\setxt$ generator, its operational reconstruction will allow for the extraction of information about the underlying dynamics. That is, the scheme presented here could be used to probe an unknown environment by observing the dynamics of the system alone (cf. Refs.~\cite{Jeske2012, NorrisPazSilva2016, probingPRA}). In fact, the full set of intermediate completely positive dynamical maps constitutes the maximum possible amount of dynamical information imprinted on the system without considering higher order multitime correlations (see Refs.~\cite{nonMarkovPRA,nonMarkovPRL}). The latter we will consider in a similar context in subsequent work.

\begin{acknowledgments}
\textbf{Acknowledgments}---The authors would like to thank Jared Cole, Guy Cohen, and C\'{e}sar Rodr\'{i}guez-Rosario for stimulating discussions regarding this work.
KM is supported through ARC FT160100073.\leavevmode\vphantom{\cite{Yang2016,jphysa}}
\end{acknowledgments}

%\begin{thebibliography}
\bibliographystyle{apsrev4-1}
\bibliography{normalNZ}
%\end{thebibliography}

%%%%%%%%%%%%%%%%%%%%%%%%%%%%%%%%%%%%
\onecolumngrid

\newpage

\appendix
%%%%%%%%%%%%%%%%%%%%%%%%%%%%%%%%%%%%

%%%%%%%%%%%%%%%%%%%%%%%%%%%%%%%%%%%%
%%%%%%%%%%%%%%%%%%%%%%%%%%%%%%%%%%%%
\section*{APPENDICES}
\section{Avoiding the inhomogeneous term}
\label{app:initialcorr}
%%%%%%%%%%%%%%%%%%%%%%%%%%%%%%%%%%%%
%%%%%%%%%%%%%%%%%%%%%%%%%%%%%%%%%%%%

Many techniques for simulating open quantum systems, exactly or approximately, rely on a product assumption for the initial state: $\rho^\se_{t_0} = \rho_{t_0}\otimes \rho^\e_{t_0}$. When this is not the case, there is always an inhomogeneous term in Eq.~\eqref{eq:memkernME}, and, while techniques have been developed to absorb it into the homogeneous part~\cite{Yang2016,jphysa}, these rely on knowledge of the underlying dynamics.  This makes the explicit calculation of $\rho_t=\mathcal{M}_{t,t_0}[\mathcal{A}]$, and hence $\Xi_{t:t_0}$, problematic for an initially correlated $\setxt$, at least when the preparation $\mathcal{A}$ does not break those correlations. 

However, when a preparation does break the correlations between system and environment, the subsequent dynamics can be simulated in the usual way (with the environment initially in state $\trs\{\mathcal{A}\rho^\se_{t_0}\}$). We now use the property that the superchannel is linear, and that any preparation operation can be written as a linear combination $\mathcal{A} = \sum_{\alpha}c_\alpha\mathcal{A}^{(\alpha)}$, where $\{\mathcal{A}^{(\alpha)}\}$ is a set of entanglement breaking completely-positive (but not necessarily trace-preserving) maps, and $c_\alpha\in \mathbb{C}$~\cite{nonMarkovPRA}. This means that the time-evolved state for any preparation can be written as a linear combination of time-evolved states from some finite set of preparations where there are no initial correlations: $\rho_{t,\mathcal{A}} = \sum_\alpha c_\alpha \rho_{t,\mathcal{A}^{(\alpha)}}$.

In the dilated picture, this is equivalent to writing the initial $\setxt$ state as $\rho_{t_0}^\se = \sum_\alpha c_\alpha X^{(\alpha)}\otimes \tau^{(\alpha)}$, where $\{X^{(\alpha)}\}$ is a set of $d_\s^2$ linearly independent system operators. By solving for the dynamics with each of the $d_\s^2$ uncorrelated initial environment states $\tau^{(\alpha)}$, the need to calculate the inhomogeneous term can be entirely circumvented.

%%%%%%%%%%%%%%%%%%%%%%%%%%%%%%%%%%%%
%%%%%%%%%%%%%%%%%%%%%%%%%%%%%%%%%%%%
\section{Simulation error}
\label{app:error}
%%%%%%%%%%%%%%%%%%%%%%%%%%%%%%%%%%%%
%%%%%%%%%%%%%%%%%%%%%%%%%%%%%%%%%%%%

Since the expression for the time-evolved density operator in Eq.~\eqref{eq:Mdecomp} is exact, errors in a simulation using transfer tensors can arise only if terms in the expansion are neglected, or otherwise from errors in the reconstruction of the maps $\Lambda_{t:s}$. Here, we assume the reconstruction is accurate, and determine the error associated with neglecting memory effects beyond some cutoff time. In what follows, we will quantify the `size' of an operator with the trace norm $\|X\|_1 = \tr\{|X|\}$, where $|X|$ is the operator formed by taking the magnitudes of the singular values of $X$ and the same singular vectors. For superoperators, we will use $\|\mathcal{X}\|:= \max_{A,B}\{\tr[A(\mathcal{X}B)]\}$, which is the largest singular value (or operator norm) of the matrix representation of $\mathcal{X}$ which acts on vectorised operators.

The error in the final state $\tilde{\rho}^{(m)}_{t_k}$ due to neglecting memory effects beyond a time $t_m = m \delta t$ is given by
\begin{align}
\left\|\rho_{t_k}-\tilde{\rho}^{(m)}_{t_k}\right\|_1 =& \left\|\sum_{j=0}^{k-1} T^{(k-j)}_{t_k:t_j}\rho_{t_j} + \Xi_{t_k:t_0} - \sum_{j=m}^{k-1} T^{(k-j)}_{t_k:t_j}\rho_{t_j}\right\|_1 = \left\|\sum_{l=m+1}^{k} T^{(l)}_{t_k:t_{k-l}}\rho_{t_{k-l}} + \Xi_{t_k:t_0} \right\|_1\nonumber\\
\leq& \sum_{l=m+1}^{k}\left\|T^{(l)}_{t_k:t_{k-l}}\right\| + \left\|\Xi_{t_k:t_0}\right\|_1 ,
\end{align}
where we have used the triangle inequality and the definition of superoperator norm given above. Referring back to Eq.~\eqref{eq:opkernel}, we can see that, for small $\delta t$ (and $t_m \gg \delta t$), 
\begin{align}
\left\|\rho_{t_k}-\tilde{\rho}^{(m)}_{t_k}\right\|_1 \lesssim & \sum_{l=m+1}^{k} \delta t^2 \|\mathcal{K}_{t_k,t_{k-l}}\|+\delta t\|\mathcal{J}_{t_k,t_0}\|_1 \simeq \delta t \left(\int_{t_m-t_0}^{t_k}\rmd s\, \|\mathcal{K}_{t_k,{t_k-s}}\|+\|\mathcal{J}_{t_k,t_0}\|_1\right).
\end{align}
That is, the error decreases with both the length of the time step $\delta t$ and the size of the tail of the memory kernel (beyond the considered memory time $t_m$). 

Importantly, under the assumption that the decay over one memory time $t_m$ (beyond the first) is significant for both $\left\|T^{(l)}_{t_k:t_{k-l}}\right\|$ and $\left\|\Xi_{t_k:t_0}\right\|_1$ (or, in the small $\delta t$ limit, $\|\mathcal{K}_{t_k,{t_k-s}}\|$ and $\|\mathcal{J}_{t_k,t_0}\|_1$), we have that 
\begin{align}
\left\|\rho_{t_k}-\tilde{\rho}^{(m)}_{t_k}\right\|_1 \leq& \sum_{l=1}^{m}\left\|T^{(2m-l)}_{(t_k-2t_m) \Mod{T}+2t_m:(t_k-2t_m) \Mod{T}+l\delta t}\right\| \nonumber \\
&\qquad\qquad+\mathcal{O}\left(\sum_{l=1}^{m}\left\|T^{(3m-l)}_{(t_k-3t_m) \Mod{T}+3t_m:(t_k-3t_m) \Mod{T}+l\delta t}\right\|\right) \nonumber \\
\simeq& \delta t \left\{\int_{0}^{t_m}\rmd s\, \left\|\mathcal{K}_{(t_k-2t_m) \Mod{T}+2t_m,{(t_k-2t_m) \Mod{T}}+s}\right\|\vphantom{+\mathcal{O}\left(\int_{0}^{t_m}\rmd s\, \left\|\mathcal{K}_{(t_k-3t_m) \Mod{T}+3t_m,{(t_k-3t_m) \Mod{T}}+s}\right\|\right)}\right.\nonumber\\
&\qquad\qquad+\left.\vphantom{\int_{0}^{t_m}\rmd s\, \left\|\mathcal{K}_{(t_k-2t_m) \Mod{T}+2t_m,{(t_k-2t_m) \Mod{T}}+s}\right\|+}\mathcal{O}\left(\int_{0}^{t_m}\rmd s\, \left\|\mathcal{K}_{(t_k-3t_m) \Mod{T}+3t_m,{(t_k-3t_m) \Mod{T}}+s}\right\|\right)\right\},
\end{align}
which at leading order does not depend on $t_k$ itself, but rather the phase with respect to the driving cycle, $(t_k-2t_m) \Mod T$. In other words, beyond a certain point, the error does not grow with the evolution time.

%%%%%%%%%%%%%%%%%%%%%%%%%%%%%%%%%%%%
%%%%%%%%%%%%%%%%%%%%%%%%%%%%%%%%%%%%
\section{Expanding transfer tensors in terms of system-environment projection superoperators}
\label{app:expTT}
%%%%%%%%%%%%%%%%%%%%%%%%%%%%%%%%%%%%
%%%%%%%%%%%%%%%%%%%%%%%%%%%%%%%%%%%%

In order to reduce notational clutter, we leave out the explicit $\setxt$ label on superoperators $\mathcal{P}$, $\mathcal{Q}$, $\mathcal{U}$ and $\mathcal{L}$ in this section. Since we are assuming underlying dynamics such that $\Lambda_{t:s}\rho=\tre\{\mathcal{U}_{t:s}\rho\otimes \tau^\e_s\}$, we can write the transfer tensors in terms of system-environment quantities; the action of the superoperator in Eq.~\eqref{eq:transfertensor} becomes
\begin{align}
T^{(N-j)}_{t:t_j}\rho =& \tre\left\{ \mathcal{U}_{t:t_j}\rho\otimes \tau_{t_j} - \sum_{k=j+1}^{N-1}\mathcal{U}_{t:t_k}\tre\left\{\mathcal{U}_{t_k:t_j} \rho \otimes \tau_{t_j} \right\}\otimes\tau_{t_k} \vphantom{+ \sum_{k=j+2}^{N-1}\sum_{l=j+1}^{k-1}\mathcal{U}_{t:t_k}\tre\left\{\mathcal{U}_{t_k:t_l} \tre\left\{\mathcal{U}_{t_l:t_j}\rho \otimes \tau_{t_j}\right\} \otimes \tau_{t_l} \right\}\otimes\tau_{t_k} -\dots}\right. \nonumber \\
& \qquad\qquad\left. \vphantom{\mathcal{U}_{t:t_j}\rho\otimes \tau_{t_j} - \sum_{k=j+1}^{N-1}\mathcal{U}_{t:t_k}\tre\left\{\mathcal{U}_{t_k:t_j} \rho \otimes \tau_{t_j} \right\}\otimes\tau_{t_k}} + \sum_{k=j+2}^{N-1}\sum_{l=j+1}^{k-1}\mathcal{U}_{t:t_k}\tre\left\{\mathcal{U}_{t_k:t_l} \tre\left\{\mathcal{U}_{t_l:t_j}\rho \otimes \tau_{t_j}\right\} \otimes \tau_{t_l} \right\}\otimes\tau_{t_k} -\dots\right\}. 
\end{align}
This can be simplified by introducing the time-dependent projection superoperators $\mathcal{P}_{t}$ and $\mathcal{Q}_t = \mathcal{I}^{\se}-\mathcal{P}_t$, defined by the action $\mathcal{P}_t X = \tre X\otimes \tau_t$. In terms of these, the transfer tensor acting on the system is
\begin{align}
T^{(N-j)}_{t:t_j}\!\rho =& \tre\left\{\mathcal{U}_{t:t_j}\mathcal{P}_{t_j}\rho^\se \!-\!\! \sum_{k=j+1}^{N-1} \mathcal{U}_{t:t_k} \mathcal{P}_{t_k} \mathcal{U}_{t_k:t_j}\mathcal{P}_{t_j}\rho^\se \!+\!\! \sum_{k=j+2}^{N-1}\sum_{l=j+1}^{k-1}\mathcal{U}_{t:t_k}\mathcal{P}_{t_k} \mathcal{U}_{t_k:t_l}\mathcal{P}_{t_l} \mathcal{U}_{t_l:t_j}\mathcal{P}_{t_j}\rho^\se \!-\! \dots\! \right\} \nonumber\\
=& \tre\left\{\mathcal{P}_{t}\mathcal{U}_{t:t_{N-1}}\mathcal{Q}_{t_{N-1}}\mathcal{U}_{t_{N-1}:t_{N-2}}\mathcal{Q}_{t_{N-2}}\dots\mathcal{Q}_{t_{j+1}}\mathcal{U}_{t_{j+1}:t_j}\mathcal{P}_{t_j}\rho^\se\right\},\label{eq:ttensorcontinuous}
\end{align}
where $\rho^\se$ is any state satisfying $\tre\{\rho^\se\}=\rho$. 
Noting that $\mathcal{M}_{t:t_0}[\mathcal{A}] = \tre\{\mathcal{U}_{t:t_0}\mathcal{A}\rho^\se_{t_0}\}$, a similar expansion for the inhomogeneous term leads to
\begin{align}
\Xi_{t:t_0} = \tre\left\{\mathcal{P}_{t}\mathcal{U}_{t:t_{N-1}}\mathcal{Q}_{t_{N-1}}\mathcal{U}_{t_{N-1}:t_{N-2}}\mathcal{Q}_{t_{N-2}}\dots\mathcal{Q}_{t_{1}}\mathcal{U}_{t_{1}:t_0}\mathcal{Q}_{t_0}\mathcal{A}\rho_{t_0}^\se\right\}. \label{eq:inhomcontinuous}
\end{align}

\begin{figure}[thp]
\centering
\includegraphics[width=0.47\textwidth]{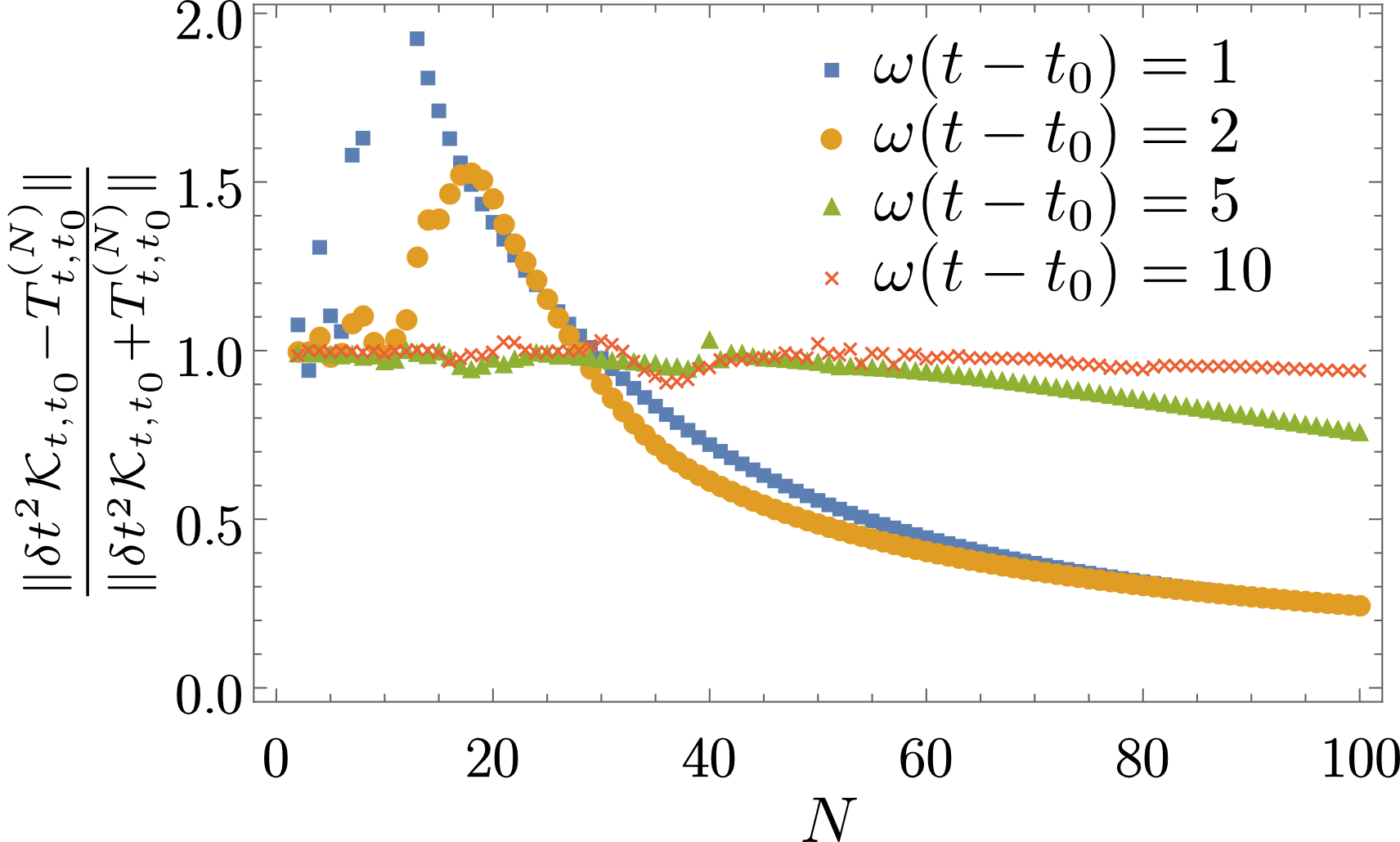}
\caption{\textbf{Emergence of the Nakajima-Zwanzig equation in the continuous limit.} Relative difference (as measured by operator norm) between scaled memory kernel $\delta t^2 \mathcal{K}_{t,t_0}$ and transfer tensor $T^{(N)}_{t,t_0}$ for different values of $t$ and $N=t/\delta t$, using the same model and parameters as in Fig.~\ref{fig:memkernexample}. Convergence is slower with $N$ for longer evolution times, though, in this case, the same $N$ corresponds to a longer $\delta t$. Note that both methods can be used to exactly calculate dynamics, even outside of the continuous limit where they converge.}
\label{fig:convergence}
\end{figure}

In the limit considered in the main text, where $t_{j+1} - t_j =\delta t := (t-t_0)/N$ and $N$ is very large, the time evolution operator between two adjacent time points can be expanded in powers of $\delta t$ as follows:
\begin{align}
\mathcal{U}_{t_{j+1}:t_j} =& T_\leftarrow \exp\left[\int_{t_j}^{t_{j+1}}\!\!\rmd s \, \mathcal{L}_s\right] %\nonumber\\
= \mathcal{I} + \int_{t_j}^{t_{j+1}}\!\!\rmd s\, \mathcal{L}_s + \int_{t_j}^{t_{j+1}}\!\!\rmd s \int_{t_j}^s\rmd s' \,\mathcal{L}_s\mathcal{L}_{s'} +\dots\nonumber\\
\simeq& \mathcal{I} + \delta t \mathcal{L}_{t_j} + \frac{1}{2}\delta t^2 \mathcal{L}_{t_j}^2 +\dots.
\end{align}
Substituting this into Eq.~\eqref{eq:ttensorcontinuous}, we have
\begin{align}
T^{(N-j)}_{t:t_j}\rho \simeq&\tre\left\{\mathcal{P}_{t}(\mathcal{I} + \delta t \mathcal{L}_{t_{N-1}} +\dots)\mathcal{Q}_{t_{N-1}}(\mathcal{I} + \delta t \mathcal{L}_{t_{N-2}} +\dots)\mathcal{Q}_{t_{N-2}}\dots\mathcal{Q}_{t_{j+1}}(\mathcal{I} + \delta t \mathcal{L}_{t_j} +\dots)\mathcal{P}_{t_j}\rho^\se\right\}\nonumber\\
=&\delta t^2\! \left(\!\tre\left\{\mathcal{P}_{t} \mathcal{L}_{t_{N-1}} \!\left(\mathcal{I} \!+\! \delta t\!\sum_{k=j+1}^{N-2}\mathcal{Q}_{t_{k}} \mathcal{L}_{t_{k}}\!+\!\delta t^2\!\sum_{k=j+2}^{N-2}\sum_{l=j+1}^{k-1}\!\mathcal{Q}_{t_{k}} \mathcal{L}_{t_{k}}\mathcal{Q}_{t_{l}} \mathcal{L}_{t_{l}} \!+\! \dots\!\right)\!\mathcal{Q}_{t_{j+1}}\mathcal{L}_{t_j}\mathcal{P}_{t_j}\rho^\se\!\right\}\right. \nonumber \\
&-\!\left. \tre\left\{\mathcal{P}_{t} \mathcal{L}_{t_{N-1}}\! \left(\mathcal{I} \!+\! \delta t\!\sum_{k=j+1}^{N-2}\!\mathcal{Q}_{t_{k}} \mathcal{L}_{t_{k}}\!+\!\delta t^2\!\sum_{k=j+2}^{N-2}\sum_{l=j+1}^{k-1}\!\mathcal{Q}_{t_{k}} \mathcal{L}_{t_{k}}\mathcal{Q}_{t_{l}} \mathcal{L}_{t_{l}} \!+\! \dots\!\right)\!\frac{\mathcal{P}_{t_{j+1}}\!-\!\mathcal{P}_{t_{j}}}{\delta t}\rho^\se\right\} \right),
\end{align}
where we have used the fact that $\mathcal{Q}_t\mathcal{Q}_s=\mathcal{Q}_s$, $\mathcal{P}_{t}\mathcal{Q}_s=0$ and $\mathcal{Q}_{t}\mathcal{P}_s= \mathcal{P}_s-\mathcal{P}_t$. In the limit that $N\rightarrow\infty$, $\delta t \rightarrow 0$ and the terms inside the central brackets resum to a time-ordered exponential (this limit is guaranteed to converge uniformly, as long as the generator is bounded and continuous~\cite{prepelita1998}): 
\begin{align}
\lim_{N\rightarrow\infty}T^{(N-j)}_{t:t_j}\rho 
=\delta t^2\, \tre\left\{\mathcal{P}_{t} \mathcal{L}_{t} T_\leftarrow \exp\left[\int_{t_j}^t\rmd s\, \mathcal{Q}_s\mathcal{L}_s\right]\left(\mathcal{Q}_{t_{j}}\mathcal{L}_{t_j}\mathcal{P}_{t_j}-\dot{\mathcal{P}}_{t_j}\right)\rho^\se\right\} = \delta t^2 \mathcal{K}_{t:t_j} \rho.
\end{align}
The convergence of this equation, for the model presented in the main text, is exemplified in Fig.~\ref{fig:convergence}. Again, the same procedure can be performed for the inhomogeneous term in Eq.~\eqref{eq:inhomcontinuous}, leading to
\begin{align}
\lim_{N\rightarrow\infty}\Xi_{t:t_0} = \delta t \,\tre\left\{\mathcal{P}_{t}\mathcal{L}_{t} T_\leftarrow \exp\left[\int_{t_j}^t\rmd s\, \mathcal{Q}_s\mathcal{L}_s\right]\mathcal{Q}_{t_0}\mathcal{A}\rho_{t_0}^\se\right\}= \delta t \mathcal{J}_{t:t_0}.
\end{align}

%%%%%%%%%%%%%%%%%%%%%%%%%%%%%%%%%%%%
%%%%%%%%%%%%%%%%%%%%%%%%%%%%%%%%%%%%
%%%%%%%%%%%%%%%%%%%%%%%%%%%%%%%%%%%%
%%%%%%%%%%%%%%%%%%%%%%%%%%%%%%%%%%%%

%%%%%%%%%%%%%%%%%%%%%%%%%%%%%%%%%%%%
%%%%%%%%%%%%%%%%%%%%%%%%%%%%%%%%%%%%
\section{Nakajima-Zwanzig with time-dependent projection operators}
\label{app:NZTDOP}
%%%%%%%%%%%%%%%%%%%%%%%%%%%%%%%%%%%%
%%%%%%%%%%%%%%%%%%%%%%%%%%%%%%%%%%%%

Here, we briefly derive the Nakajima Zwanzig equation given by Eqs.~\eqref{eq:memkernME}~\&~\eqref{eq:memkern} from the underlying equation of motion given in Eq.~\eqref{eq:underlying}. Starting from the definition of the projection superoperators $\mathcal{P}^\se_t$ and $\mathcal{Q}^\se_t = \mathcal{I}^\se-\mathcal{P}^\se_t$ (where $\mathcal{P}^\se_t X^\se = \tre\{X^\se\}\otimes \tau_t^\e$), we first note the following identities:
\begin{align}
\mathcal{P}^\se_t \mathcal{P}^\se_s =& \tre\left\{\tre\{\,\cdot\,\}\otimes \tau^\e_s\right\}\otimes\tau^\e_t = \tre\{\,\cdot \,\}\otimes \tau^\e_t = \mathcal{P}^\se_t,\\
\mathcal{Q}^\se_t \mathcal{Q}^\se_s =& %\left(\mathcal{I}^\se-\mathcal{P}^\se_t\right)\left(\mathcal{I}^\se-\mathcal{P}^\se_s\right) =
\mathcal{I}^\se-\mathcal{P}^\se_t-\mathcal{P}^\se_s+\mathcal{P}^\se_t\mathcal{P}^\se_s = \mathcal{I}^\se-\mathcal{P}^\se_s = \mathcal{Q}^\se_s,\\
\mathcal{Q}^\se_t \mathcal{P}^\se_s =& \mathcal{P}^\se_s-\mathcal{P}^\se_t\mathcal{P}^\se_s = \mathcal{P}^\se_s-\mathcal{P}^\se_t,\\
\mathcal{P}^\se_t \mathcal{Q}^\se_s =& \mathcal{P}^\se_t-\mathcal{P}^\se_t\mathcal{P}^\se_s =0.
\end{align}

Next, we consider the equations of motion of the so-called `relevant' and `irrelevant' parts of the $\setxt$ density operator, $\mathcal{P}^\se_t\rho^\se_t$ and $\mathcal{Q}^\se_t\rho^\se_t$ respectively. Using Eq.~\eqref{eq:underlying}, we find
\begin{gather}
\frac{\rmd}{\rmd t} \mathcal{P}^\se_t\rho^\se_t = (\mathcal{P}^\se_t\mathcal{L}^\se_t\mathcal{P}^\se_t+\dot{\mathcal{P}}^\se_t)\rho^\se_t + \mathcal{P}^\se_t\mathcal{L}^\se_t\mathcal{Q}^\se_t\rho^\se_t, \label{eq:relpart}
\end{gather}
and
\begin{gather}
\frac{\rmd}{\rmd t} \mathcal{Q}^\se_t\rho^\se_t = (\mathcal{Q}^\se_t\mathcal{L}^\se_t\mathcal{P}^\se_t-\dot{\mathcal{P}}^\se_t)\rho^\se_t + \mathcal{Q}^\se_t\mathcal{L}^\se_t\mathcal{Q}^\se_t\rho^\se_t, \label{eq:irrelpart}
\end{gather}
where we have used that $\mathcal{I}^\se =\mathcal{P}^\se_t+ \mathcal{Q}^\se_t$. Formally solving Eq.~\eqref{eq:irrelpart} gives
\begin{gather}
\mathcal{Q}^\se_t\!\rho^\se_t \!=\! T_\leftarrow \exp\!\left[\int_{t_0}^t\!\rmd s\, \mathcal{Q}^\se_s\mathcal{L}^\se_s\right]\mathcal{Q}^\se_{t_0}\mathcal{A}\rho_{t_0}^\se \!+\! \int_{t_0}^t \!\rmd s\, T_\leftarrow \exp\!\left[\int_{s}^t\rmd s'\, \mathcal{Q}^\se_{s'}\mathcal{L}^\se_{s'}\right] \!(\mathcal{Q}^\se_s\mathcal{L}^\se_s\mathcal{P}^\se_s\!-\!\dot{\mathcal{P}}^\se_s)\rho^\se_s.
\end{gather}
As in the main text, we have taken the post-preparation $\setxt$ state $\mathcal{A}\rho^\se_{t_0}$ as our initial condition. Substituting this into Eq.~\eqref{eq:relpart} and taking the partial trace over $\etxt$ (since $\dot{\rho}_t = \tre\left\{\rmd/\rmd t \mathcal{P}^\se_t\rho^\se_t\right\}$) leads to
\begin{align} \label{eq:NZderived}
\dot{\rho}_t =& \tre\left\{\mathcal{P}^\se_t\mathcal{L}^\se_t\mathcal{P}^\se_t\rho^\se_t\right\} \nonumber \\
&+  \tre\left\{\mathcal{P}^\se_t \mathcal{L}^\se_t T_\leftarrow\!\exp\left[\int_{t_0}^t\rmd s\, \mathcal{Q}^\se_s \mathcal{L}^\se_s\right]\mathcal{Q}^\se_{t_0}\mathcal{A}\rho^\se_{t_0}\right\}\nonumber \\
&+ \int_{t_0}^t\rmd  s\,\tre\left\{\mathcal{P}^\se_t \mathcal{L}^\se_t T_\leftarrow\exp\left(\int_{s}^t\rmd  s'\, \mathcal{Q}^\se_{s'} \mathcal{L}^\se_{s'}\right)\left(\mathcal{Q}^\se_{s} \mathcal{L}^\se_{s} \mathcal{P}^\se_{s}-\dot{\mathcal{P}}^\se_{s} \right) 
\rho^\se_{s}\right\}.
\end{align}
In the first line, the derivative-dependent term vanishes, since $\tr\{\dot{\tau}^\e_t\} = 0$, and hence $\tre \{\dot{\mathcal{P}}^\se_t\rho^\se_t\} = 0$, when $\tau^\e_t$ is constrained to be a unit trace density operator (as we will always assume). Since $\mathcal{P}^\se_t \rho^\se_t = \mathcal{P}^\se_t \rho_t\otimes x^\e$ and $\dot{\mathcal{P}}^\se_t \rho^\se_t = \dot{\mathcal{P}}^\se_t \rho_t\otimes x^\e$ for any unit trace operator $x^\e$, we can identify each line of Eq.~\eqref{eq:NZderived} with one of the terms in Eqs.~\eqref{eq:memkern}.

Our equation differs from that in Refs.~\cite{WillisPicard1974,PicardWillis1977}, since we do not assume that $\dot{\mathcal{P}}^\se_{t}\rho^\se_t = 0$~$\forall t$. If we were to do so, then the derivative term in the last line of Eq.~\eqref{eq:NZderived} would vanish, and we would recover the form derived in the aforementioned references.

\end{document}